\documentclass{aa}  

\usepackage{graphicx}
\usepackage{txfonts}
\usepackage{lipsum}
\usepackage{subcaption}         
\usepackage{lscape}             
\usepackage{placeins}           
                                
\begin{document}

   \title{HARPS-N Reveals a Well-aligned Orbit for the Highly Eccentric Warm Jupiter TOI-4127~b}

   \subtitle{}

   \author{I. Mireles\inst{1}, F. Murgas\inst{2, 3}, D. Dragomir\inst{1}, E. Palle\inst{2, 3}, J. Dong\inst{4, 5}, I. Carleo\inst{6} \and E. Esparza-Borges\inst{2, 3}}

   \institute{Department of Physics and Astronomy, The University of New Mexico, Albuquerque, NM 87106, USA\\
    \email{mirelesi@unm.edu}
    \and Instituto de Astrofísica de Canarias (IAC), 38200 La Laguna, Tenerife, Spain
    \and Departamento de Astrofísica, Universidad de La Laguna (ULL), 38206 La Laguna, Tenerife, Spain
    \and Center for Computational Astrophysics, Flatiron Institute, 162 Fifth Avenue, New York, NY 10010, USA
    \and Department of Astronomy, University of Illinois at Urbana-Champaign, Urbana, IL 61801, USA
    \and INAF -- Osservatorio Astrofisico di Torino, Via Osservatorio 20, I-10025, Pino Torinese, Italy}

   \date{Received September 30, 20XX}

 
  \abstract
   {While many hot Jupiter systems have a measured obliquity, few warm Jupiter systems do. The longer orbital periods and transit durations of warm Jupiters make it more difficult to measure the obliquities of their host stars. However, the longer periods also mean any misalignments persist due to the longer tidal realignment timescales. As a result, measuring these obliquities is necessary to understand how these types of planets form and how their formation and evolution differ from that of hot Jupiters.}
   {Here, we report the measurement of the Rossiter-McLaughlin effect for the TOI-4127 system using the HARPS-N spectrograph.} 
   {We model the system using our new HARPS-N radial velocity measurements in addition to archival TESS photometry and NEID and SOPHIE radial velocities.}
   {We find that the host star is well-aligned with the highly eccentric (e=0.75) warm Jupiter TOI-4127~b, with a sky-projected obliquity $\lambda = 4_{-16}^{+17}\, ^{\mathrm{\circ}}$. This makes TOI-4127 one of the most eccentric well-aligned systems to date and one of the longest period systems with a measured obliquity.}
   {The origin of its highly eccentric yet well-aligned orbit remains a mystery, however, and we investigate possible scenarios that could explain it. While typical in-situ formation and disk migration scenarios cannot explain this system, certain scenarios involving resonant interactions between the planet and protoplanetary disc could. Similarly, specific cases of planet-planet scattering or Kozai-Lidov oscillations can result in a highly-eccentric and well-aligned orbit. Coplanar high-eccentricity migration could also explain this system. However, both this mechanism and Kozai-Lidov oscillations require an additional planet in the system that has not yet been detected.}


   \titlerunning{TOI-4127 RM Effect}
   \authorrunning{I. Mireles et al.}
   \maketitle

\section{Introduction} \label{sec:intro}
Despite their intrinsic rarity, close-in giant planets make up a significant portion of the over 6000 exoplanets discovered to date. These close-in giants can be divided into two populations known as hot and warm Jupiters which have significant differences in system and orbital architectures. Warm Jupiters are located far enough from their host stars that tidal dissipation is not efficient at circularizing or realigning their orbits, with pericenter distances a(1-e) $\succsim$ 0.1 AU. For a warm Jupiter in a  circular orbit around a Sun-like star, this corresponds to an orbital period of $\sim$10 days. While determining the eccentricities of these planets comes naturally from also measuring the masses from RVs, measuring the spin-orbit alignment of these systems is more challenging. The long orbital periods and transit durations of warm Jupiters results in fewer opportunities to measure the stellar obliquity through the Rossiter-McLaughlin (RM) effect \citep{1924ApJ....60...22M, 1924ApJ....60...15R}. The RM effect can only measure the sky-projected stellar obliquity, $\lambda$, as obtaining the true 3-dimensional obliquity, $\psi$, requires knowing the stellar inclination. However, measuring the inclination of a host star requires asteroseismology or a precise rotation period, which are not always readily obtainable.

Although many hot Jupiter host stars have measured obliquities, very few warm Jupiter host stars do \citep{2022PASP..134h2001A}. Hot Jupiter systems show a wide range of obliquities \citep{2023AJ....166..112D, 2023ApJ...950L...2S}, especially hot Jupiters around stars above the Kraft break \citep{1967ApJ...150..551K} that show much greater misalignment than ones orbiting cooler stars \citep{2010ApJ...718L.145W}. On the other hand, the few warm Jupiter systems with obliquity measurements generally show well-aligned systems regardless of host stars \citep{2022AJ....164..104R, 2024ApJ...973L..21W}. Unlike the case of many well-aligned hot Jupiters, tidal realignment cannot explain well-aligned warm Jupiters as the realignment timescales exceed the lifetimes of the host stars even in the case of cool dwarfs \citep{2012ApJ...757...18A}. As such, the impact of a system's dynamical history is preserved.
Of the warm Jupiter systems with significantly misaligned orbits, most contain a stellar companion \citep{2023ApJ...958L..20E}.

In addition to having more generally well-aligned orbits, warm Jupiter systems are also more likely to host nearby additional planets than hot Jupiter systems \citep{2016ApJ...825...98H, 2023AJ....165..171W}, indicating that different formation and evolution pathways are responsible for the two types of systems. Warm Jupiter systems are thought to largely be the product of in-situ formation \citep[e.g.,][]{2016ApJ...817L..17B} or disk migration \citep[e.g.,][]{2014prpl.conf..667B}, both of which allow for nearby planets and also produce well-aligned orbits. On the other hand, hot Jupiter systems are likely the result of high-eccentricity migration mechanisms \citep{2016ApJ...829..132P}, processes that result in isolated planets often in misaligned orbits. Such mechanisms include Kozai-Lidov oscillations \citep[e.g.,][]{2007ApJ...669.1298F} and planet-planet scattering \citep[e.g.,][]{2014ApJ...786..101P}, although the latter mechanism can occasionally result in well-aligned short period giant planets \citep{2002Icar..156..570M}. As the orbits of migrating hot Jupiters shrink and circularize, there will be a time when they will have orbital periods the same as those of warm Jupiters. If the migration mechanism is not efficient, they can end up ``stuck" at these longer orbital periods in highly-eccentric orbits. As such, it is possible that some of the warm Jupiters observed now are actually either hot Jupiters migrating to their final orbit or ``failed" hot Jupiters. If some of the more eccentric warm Jupiters are indeed migrating or failed hot Jupiters, then they should likely have misaligned orbits. This is thought to be the case for extremely-eccentric and misaligned warm Jupiters such as HD 80606\,b \citep{2009A&A...502..695P} and the recently discovered TIC 241249530\,b \citep{2024Natur.632...50G}. However, these highly eccentric warm Jupiters are not always misaligned, as is the case with TOI-3362\,b \citep{2023ApJ...958L..20E}.

Here, we present the Rossiter-McLaughlin effect measurement for the TOI-4127 system. TOI-4127~b is a highly eccentric (e=0.75) warm Jupiter orbiting a F star ($R_{\star}$ = 1.29 $R_{\odot}$, $T_{\rm eff}$ = 6096 K) on a 56.4 day orbit \citep{2023AJ....165..234G}. Using new HARPS-N radial velocity measurements alongside archival TESS photometry and NEID and SOPHIE spectroscopy, we obtain a sky-projected obliquity of $4_{-16}^{+17}\, ^{\mathrm{\circ}}$. This makes TOI-4127 the newest addition to the small but growing sample of highly-eccentric yet well-aligned warm Jupiter systems. In Section \ref{sec:observations}, we describe the newly obtained HARPS-N spectroscopy and archival TESS photometry and NEID and SOPHIE spectroscopy used to model the system. We detail our analysis to characterize the orbit and obtain system parameters in Section \ref{sec:analysis}. We discuss possible formation and evolution scenarios and the system's place in the larger sample of warm Jupiter systems in Section \ref{sec:discussion}. We summarize our results in Section \ref{sec:conclusion}.

\section{Observations} \label{sec:observations}

\subsection{HARPS-N Spectroscopy}

We observed a transit of TOI-4127\,b on 15 March 2024 using the High Accuracy Radial velocity Planet Searcher North (HARPS-N) spectrograph at the Telescopio Nazionale Galileo (TNG) at the Roque de los Muchachos Observatory, which observes at a wavelength range of 378-691 nm with a resolution R=115,000 \citep{2012SPIE.8446E..1VC}. We obtained 15 1200-second observations spanning 4.8 hours covering a 1.7 hours of pre-transit baseline and 3.1 hours of the transit. We also obtained one 1200-second pre-transit observation on 24 November 2023 with HARPS-N before high humidity prevented the rest of the observations from being taken. The airmasses ranged from 1.38 to 1.8, while the signal-to-noise ratio ranged from 26.8 to 37.1. We obtain a median RV uncertainty of 6.88 m s$^{-1}$ across the 16 observations.

The HARPS-N spectra time series were reduced using the offline data reduction software (DRS), which also computes radial velocities based on the cross-correlation function (CCF) method \citep{Pepe2002}. In this method, the spectra are cross-correlated with a binary mask created from a template spectrum matching the target star's type. For our analysis, we chose the G2 mask, as it was the closest available match to the stellar type of TOI-4127. We reprocessed the data using the Yabi web application \citealt{yabi}, accessible at the IA2 Data Centre\footnote{\url{https://www.ia2.inaf.it/}}. The RVs are presented in Table \ref{table:harpsn_vals} of the Appendix.

\begin{figure}
\includegraphics[width=\columnwidth]{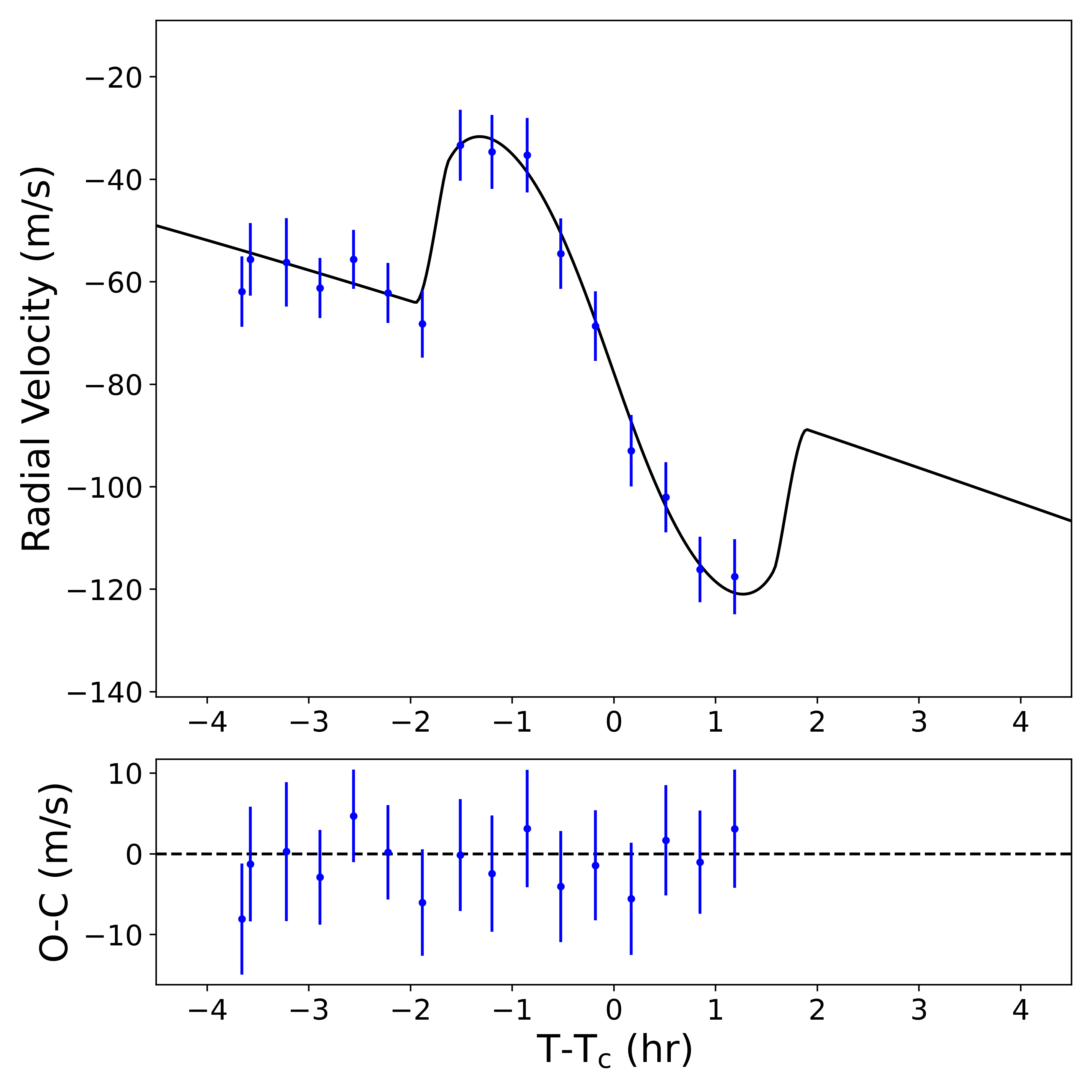}
  \caption{Phase-folded HARPS-N radial velocities and best-fit model.}
\label{fig:harpsn_rm}
\end{figure}

\subsection{Archival Observations}

\begin{figure*}
\includegraphics[width=\textwidth]{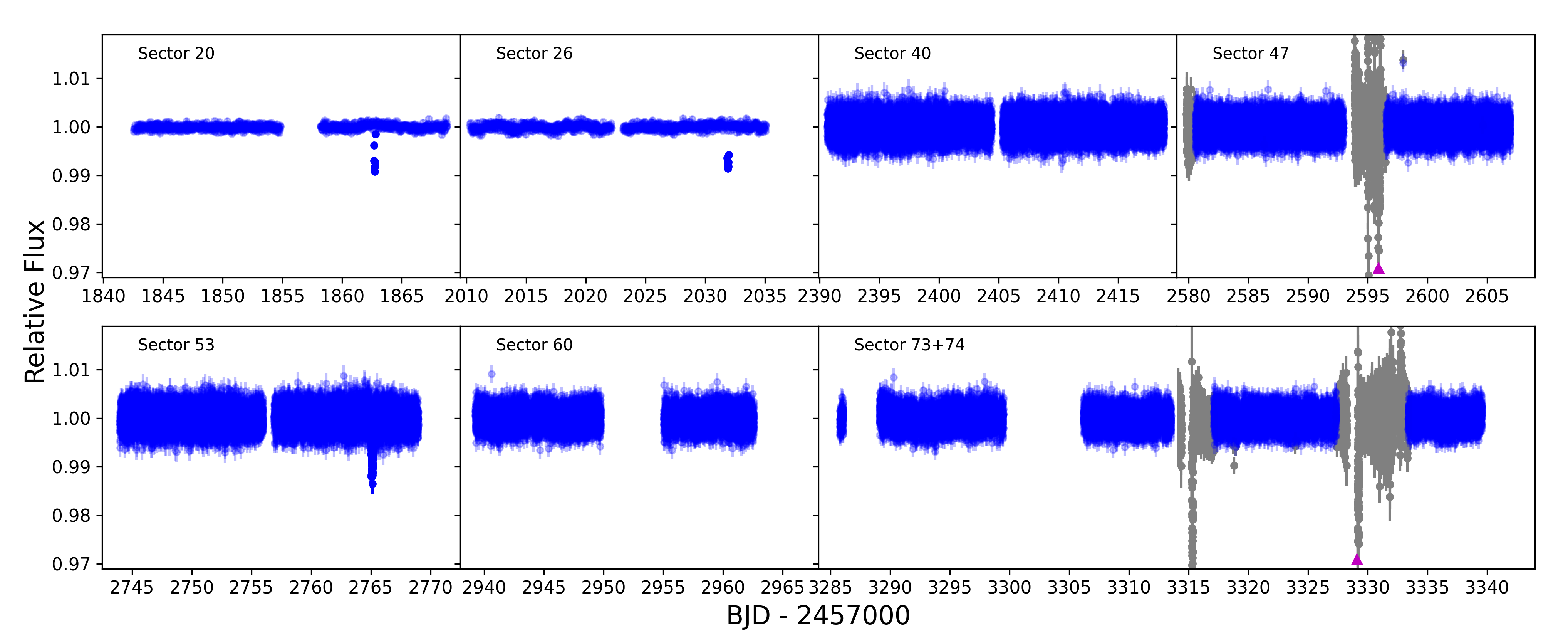}
  \caption{TESS PDCSAP 30-minute (sectors 20 and 26) and 2-minute (sectors 40, 47, 53, 60, 73, and 74) cadence photometry of TOI-4127. Gray points denote low-quality Simple Aperture Photometry (SAP) photometry not included in the analysis, while the magenta triangles denote the transit times of the two transits contaminated by scattered light.}
\label{fig:tess_lc_full}
\end{figure*}
\begin{figure}
\includegraphics[width=\columnwidth]{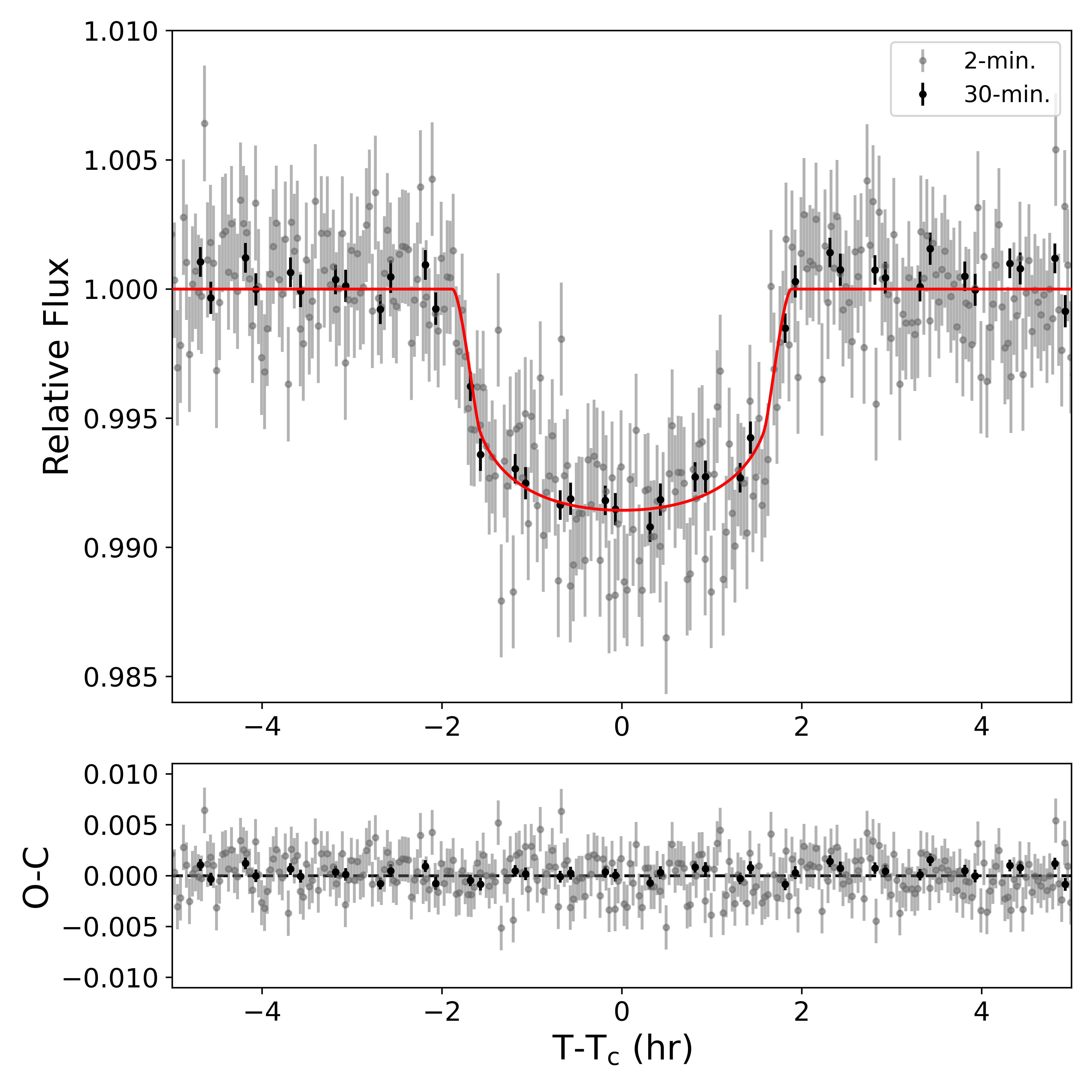}
  \caption{Phase-folded \textit{TESS} light curve zoomed in on the transit. The black and gray points denote 30-minute and 2-minute cadence data, respectively. The best-fit model is shown in red.}
\label{fig:tess_lc}
\end{figure}

We use the \textit{TESS} \citep{2015JATIS...1a4003R} Presearch Data Conditioning Simple Aperture Photometry \citep[PDCSAP;][]{2012PASP..124.1000S, 2012PASP..124..985S, 2014PASP..126..100S} photometry used by \citet{2023AJ....165..234G}. This consists of two sectors (20 and 26) of 30-minute cadence full-frame image data and three sectors (40, 47, and 53) of 2-minute cadence postage stamps. TOI-4127~b transited in 4 of the 5 sectors, however the transit in sector 47 was contaminated by scattered light and not included in the analysis (Figure \ref{fig:tess_lc_full}). Although the target was re-observed in 2-minute cadence in sectors 73 and 74, the transit in sector 74 was also contaminated by scattered light and thus also not included in the analysis. The three uncontaminated transits do not show any evidence for significant transit timing variations \citep{2023AJ....165..234G} and are shown alongside the best-fit model in Figure \ref{fig:tess_lc}.

\begin{figure}
\includegraphics[width=\columnwidth]{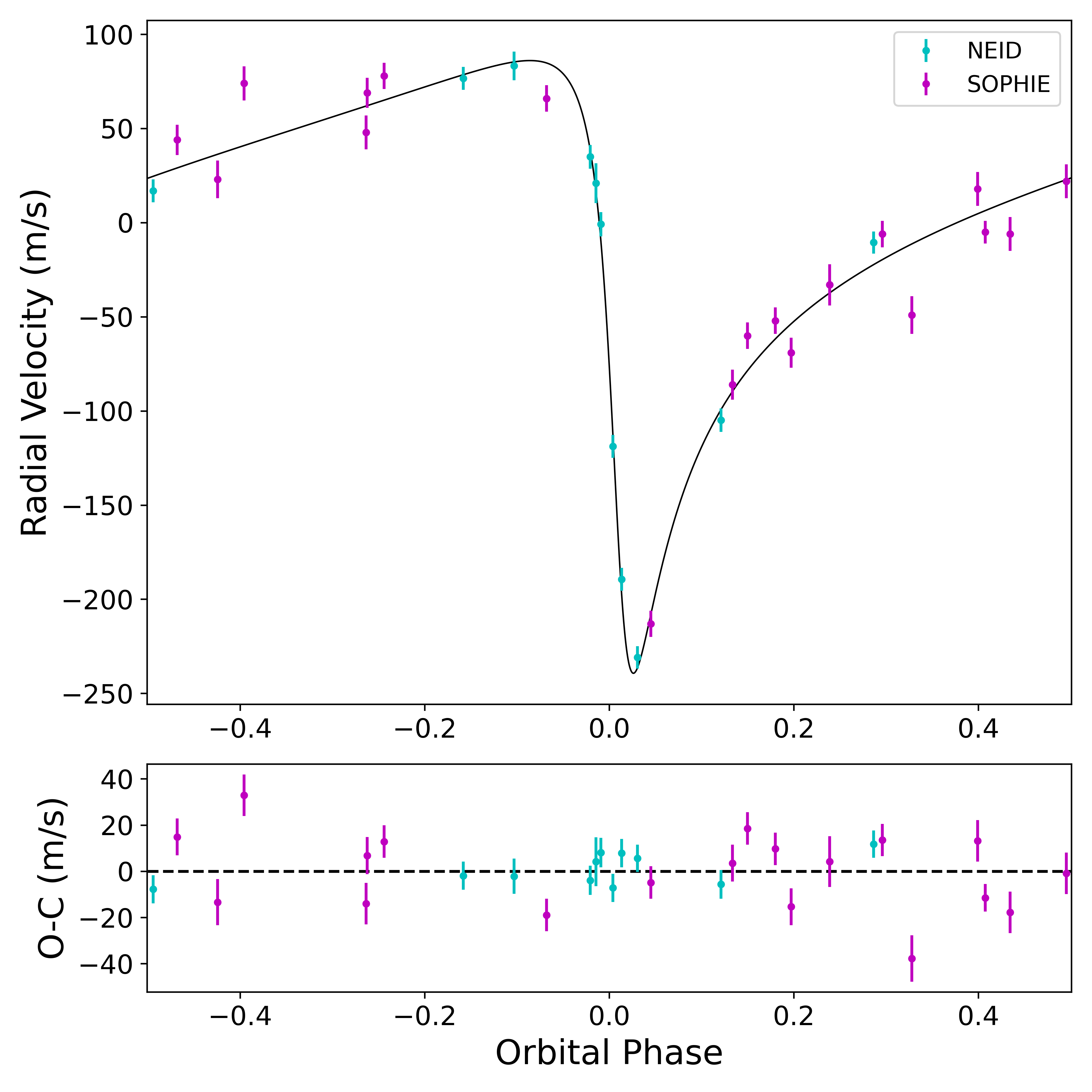}
  \caption{Phase-folded NEID (cyan) and SOPHIE (magenta) radial velocities, with the best-fit model shown in black.}
\label{fig:neid_sophie}
\end{figure}

We also include the 30 radial velocity measurements used by \citet{2023AJ....165..234G} in our analysis -- 11 from the NEID spectrograph on the WIYN 3.5m telescope at Kitt Peak National Observatory \citep{2016SPIE.9908E..7HS} and 19 from the SOPHIE spectrograph on the 1.93 m telescope at the Observatoire de Haute-Provence, France \citep{2008SPIE.7014E..0JP}. The data and best-fit model are shown in Figure \ref{fig:neid_sophie}.

We use archival high-resolution imaging from NESSI on the WIYN 3.5m telescope to help constrain possible unseen stellar companions, as described in Section \ref{subsec:comps}.

\section{Analysis} \label{sec:analysis}

\subsection{Stellar Parameters}
We adopt the stellar parameters from \citet{2023AJ....165..234G}, which are shown in Table \ref{table:system_params}. The spectroscopic parameters were derived by using \texttt{SpecMatch-Emp} \citep{2017ApJ...836...77Y} to compare the NEID spectra to reference spectra, while the physical parameters were derived through SED analysis and mass-radius relations.
We attempted to determine the stellar inclination by measuring the rotation period of the host star using a Generalized Lomb-Scargle (GLS) periodogram on the TESS data. While some individual sectors of PDCSAP data showed statistically significant peaks, the locations of the peaks varied sector by sector and were non-existent in the SAP data as well as full frame image data from the Quick-Look Pipeline \citep[QLP;][]{2020RNAAS...4..204H,2020RNAAS...4..206H}. As such, we are unable to obtain a rotation period and therefore stellar inclination, meaning we cannot yet obtain the true 3-D obliquity.

\begin{table}
\caption{System Information}
\label{table:system_params}
\centering
\renewcommand{\arraystretch}{1.25}
\begin{tabular}{l l l}
\hline\hline
Parameter & Value & Source \\
\hline
TIC & 141488193 &  TIC V8$^{a}$ \\
TOI & 4127 & TIC V8 \\
R.A. &  07:01:37.09 & ExoFOP \\
Dec. & +72:24:53.75 & ExoFOP \\
$\mu_{ra}$ (mas yr$^{-1}$)  & $3.183 \pm 0.016$ & Gaia DR3$^{b}$\\
$\mu_{dec}$ (mas yr$^{-1}$) & $-3.139 \pm 0.024$ & Gaia DR3\\
Parallax (mas) & $3.074 \pm 0.020$ & Gaia DR3 \\
$B_T$ (mag)    & $12.02 \pm 0.15$ & Tycho-2 \\
$V_T$ (mag)    & $11.44\pm 0.01$ & Tycho-2 \\
$Gaia$ (mag) &  $11.409 \pm 0.003$ & Gaia DR3 \\
$B_P$ (mag) &  $11.684 \pm 0.003$ &  Gaia DR3 \\
$R_P$ (mag) & $10.981 \pm 0.004$ &  Gaia DR3 \\
TESS (mag) & $11.04 \pm 0.01$ & TIC V8\\
$J$ (mag)    & $10.538 \pm 0.023$ & 2MASS$^{c}$  \\
$H$ (mag)    & 	$10.308 \pm 0.023$ & 2MASS \\
$K_S$ (mag)  & $10.245 \pm 0.016$ & 2MASS  \\
$T_{\rm eff}$ (K)   & $6096 \pm 115 $ & \citet{2023AJ....165..234G} \\ 
$[$Fe/H$]$  & $0.14 \pm 0.12$  & Gupta et al. (2023) \\
$\log g$  & $4.26 \pm 0.14$ & Gupta et al. (2023) \\
$M_{\star}$ ($M_{\odot}$)  & $1.230 \pm 0.070$ & Gupta et al. (2023) \\
$R_{\star}$ ($R_{\odot}$)  & $1.293 \pm 0.050$ & Gupta et al. (2023) \\ 
$L_{\star, \mathrm{bol}}$ ($L_{\odot}$)   & $2.072 \pm 0.028$ & Gupta et al. (2023) \\ 
Age (Gyr) &  $4.8 \pm 2.1$ & Gupta et al. (2023) \\ 
\hline
\multicolumn{3}{l}{Note: (a) \cite{2018AJ....156..102S}.} \\
\multicolumn{3}{l}{(b) \cite{2023AA...674A...1G}.} \\
\multicolumn{3}{l}{(c) \cite{2003yCat.2246....0C}}
\end{tabular}
\end{table}

\subsection{Joint Photometric-Spectroscopic Analysis}
We fit for system parameters using a modified version of the \texttt{allesfitter} package \citep{2021ApJS..254...13G, allesfitter-code} to perform a joint analysis of the \textit{TESS} photometry and NEID, SOPHIE, and HARPS-N radial velocities. The original version of \texttt{allesfitter} employs the \texttt{ellc} package \citep{ellc} to model the light curve and RVs. \texttt{ellc} uses the formulation from \cite{Ohta2005} to model the RM effect, which does not include microturbulence or macroturbulence. The modified version we use was developed to include these parameters \citep{Wang2024}, and uses the formulation from \cite{Hirano2011} to model the RM effect through the \texttt{tracit} package \citep{Hjorth2021, KnudstrupAlbrecht2022}. The light curve and out-of-transit RVs are modeled using the \texttt{PyTransit} \citep{Parviainen2015} and \texttt{RadVel} \citep{Fulton2018RadVel} packages, respectively.   We used a nested sampling algorithm consisting of 500 live points to explore the parameter space and determine the best-fit values for the parameters listed in Table \ref{table:planetary_params}, using uniform priors for all except two parameters.
We use empirical relations from \cite{Doyle2014} and \cite{Bruntt2010} to determine the values of the macroturbulence and microturbulence, respectively. We then use normal priors centered on those values for the two parameters, with $\sigma=1$ km s$^{-1}$.

\begin{table*}
\caption{Planetary Parameters}
\label{table:planetary_params}
\centering
\renewcommand{\arraystretch}{1.25}
\begin{tabular}{l l l l}
\hline\hline
Parameter & Description & Prior & Value \\
\hline
\textit{Modeled Parameters} & & & \\
$q_{1;TESS}$ & Linear limb-darkening coefficient for TESS & $U(0,1)$ & $0.23_{-0.08}^{+0.17}$ \\ 
$q_{2;TESS}$ & Quadratic limb-darkening coefficient for TESS & $U(0,1)$ & $0.66_{-0.28}^{+0.23}$ \\
$q_{1;HARPS-N}$ & Linear limb-darkening coefficient for HARPS-N & $U(0,1)$ & $0.52_{-0.33}^{+0.32}$ \\ 
$q_{2;HARPS-N}$ & Quadratic limb-darkening coefficient for HARPS-N & $U(0,1)$ & $0.43_{-0.26}^{+0.30}$ \\
$(R_{\star} + R_p) / a$ & Sum of radii over semi-major axis & $U(0,1)$ & $0.0215_{-0.0005}^{+0.0007}$\\ 
$R_p / R_{\star}$ & Planet-to-star radius ratio & $U(0,1)$ & $0.086 \pm 0.001$\\ 
$\cos{i}$ & Cosine of the orbital inclination & $U(0,1)$ & $0.0106_{-0.0073}^{+0.0111}$ \\ 
$T_{0}$ & Transit epoch - 2457000 (BJD) & $U(2765.0088,2765.2088)$ & $2765.1094 \pm 0.0008$ \\
$P$ & Orbital period (d) & $U(56.2,56.6)$ & $56.3987 \pm 0.0001$ \\ 
$K$ & Radial velocity semi-amplitude (m s$^{-1}$) & $U(0,1000)$ & $159.6_{-3.1}^{+3.0}$ \\
$\sqrt{e}\cos{\omega}$ & & $U(-1,1)$ & $-0.544 \pm 0.019$  \\ 
$\sqrt{e}\sin{\omega}$ & & $U(-1,1)$ & $0.673_{-0.016}^{+0.015}$ \\
$\lambda$ & Sky-projected spin-orbit angle (deg) & $U(-180,180)$ & $4_{-16}^{+17}$ \\
$v\, sin\, i$ & Host star projected rotational velocity (km s$^{-1}$) & $U(0, 20)$ & $6.30_{-0.53}^{+0.54}$ \\
$\xi$ & Microturbulent velocity (km s$^{-1}$) & $N(1.18, 1)$ & $1.29_{-0.76}^{+0.88}$ \\
$\zeta$ & Macroturbulent velocity (km s$^{-1}$) & $N(4.52, 1)$ & $4.55_{-0.96}^{+0.95}$ \\
$\ln{\sigma_\mathrm{TESS}}$ & Jitter term for 2-min TESS data & $U(-23, 0)$ & $-6.151 \pm 0.025$\\ 
$\ln{\sigma_\mathrm{TESS;LC}}$ & Jitter term for 30-min TESS data & $U(-23, 0)$ & $-7.596_{-0.070}^{+0.073}$ \\ 
$\ln{\sigma_\mathrm{jitter, SOPHIE}}$ & Jitter term for SOPHIE (ln km s$^{-1}$) & $U(-12, 0)$ & $-4.23_{-0.23}^{+0.22}$ \\ 
$\gamma_\mathrm{SOPHIE}$ & Offset term for SOPHIE (m s$^{-1}$) & $U(-1,1)$ & $0.0_{-3.8}^{+3.6}$ \\ 
$\ln{\sigma_\mathrm{jitter, NEID}}$ & Jitter term for NEID (ln km s$^{-1}$) & $U(-12, 0)$ & $-6.94_{-3.32}^{+1.70}$\\ 
$\gamma_\mathrm{NEID}$ & Offset term for NEID (m s$^{-1}$) & $U(-1,1)$ & $-0.2_{-2.8}^{+2.9}$  \\ 
$\ln{\sigma_\mathrm{jitter, HARPS-N}}$& Jitter term for HARPS-N (ln km s$^{-1}$) & $U(-12, 0)$ & $-10.11_{-3.16}^{+2.82}$ \\ 
$\gamma_\mathrm{HARPS-N}$ & Offset term for HARPS-N (m s$^{-1}$) & $U(-37000,-36000)$ & $-36764.2_{-3.7}^{+3.8}$ \\ 
 & & & \\
\textit{Derived Parameters} & & & \\
$a/R_{\star}$ & Semi-major axis over stellar radius & & $50.5_{-1.6}^{+1.2}$\\ 
$a$ & Semi-major axis (AU) & & $0.303_{-0.015}^{+0.014}$\\ 
$R_\mathrm{p}$ & Planet radius ($R_\mathrm{\oplus}$) & & $12.07_{-0.49}^{+0.50}$ \\ 
$R_\mathrm{p}$ & Planet radius ($R_\mathrm{J}$) & & $1.08 \pm 0.04$ \\ 
$M_\mathrm{p}$ & Planet mass ($M_\mathrm{J}$) & & $2.34_{-0.17}^{+0.19}$ \\
$i$ & Orbital inclination (deg) & & $89.39_{-0.64}^{+0.42}$ \\ 
$e$ & Orbital eccentricity & & $0.7484_{-0.0068}^{+0.0064}$ \\
$\omega$ & Argument of periastron (deg) & & $128.9 \pm 1.6$ \\
$b$ & Impact parameter & & $0.15_{-0.10}^{+0.15}$ \\ 
$T_\mathrm{tot}$ & Total transit duration (h) & & $3.82_{-0.04}^{+0.05}$ \\ 
$T_\mathrm{full}$ & Full transit duration (h) & & $3.20 \pm 0.04$ \\ 
$\rho_\mathrm{\star}$ & Derived stellar density (g cm$^{-3}$) & & $0.765_{-0.071}^{+0.057}$\\ 
$T_\mathrm{eq}$ & Planet equilibrium temperature$^{a}$ (K) & & $556_{-13}^{+14}$ \\
\hline
\multicolumn{3}{l}{Note: (a) Assuming an albedo of 0.3 and emissivity of 1.}
\end{tabular}
\end{table*}

The values and uncertainties of the fitted and derived parameters listed in Table \ref{table:planetary_params} are defined as the median values and 68\% confidence intervals of the posterior distributions, respectively. The best-fit phase folded transit model is shown alongside the \textit{TESS} data in Figure \ref{fig:tess_lc}. The best-fit RV model and NEID and SOPHIE data are shown in Figure \ref{fig:neid_sophie} while the best-fit RM model and HARPS-N data are shown in Figure \ref{fig:harpsn_rm}. The corner plots for the modeled and derived parameters are shown in Figure \ref{fig:corner_fitted} and Figure \ref{fig:corner_derived} of the Appendix. 

\section{Discussion} \label{sec:discussion}

\begin{figure*}
\includegraphics[width=0.49 \textwidth]{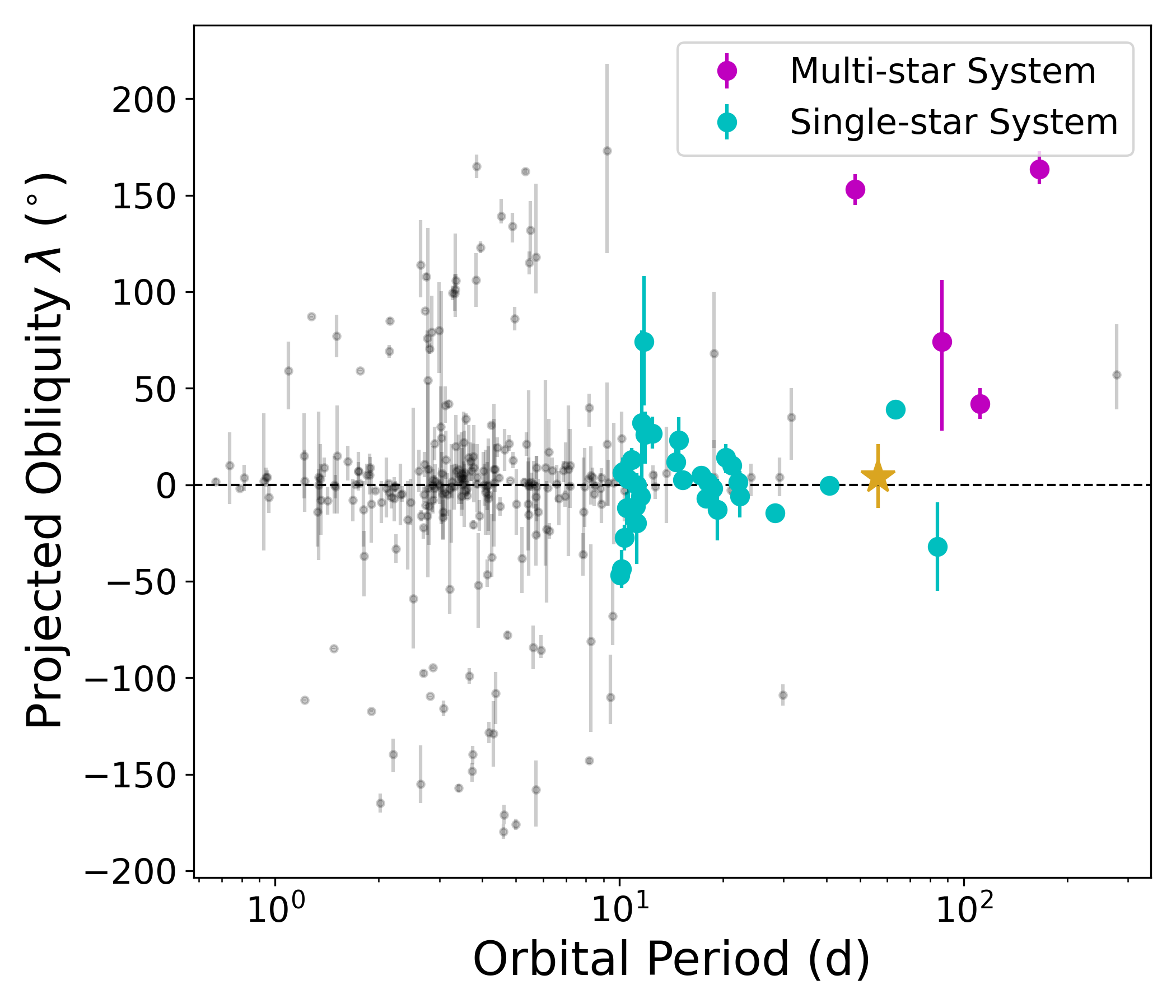}
\includegraphics[width=0.49 \textwidth]{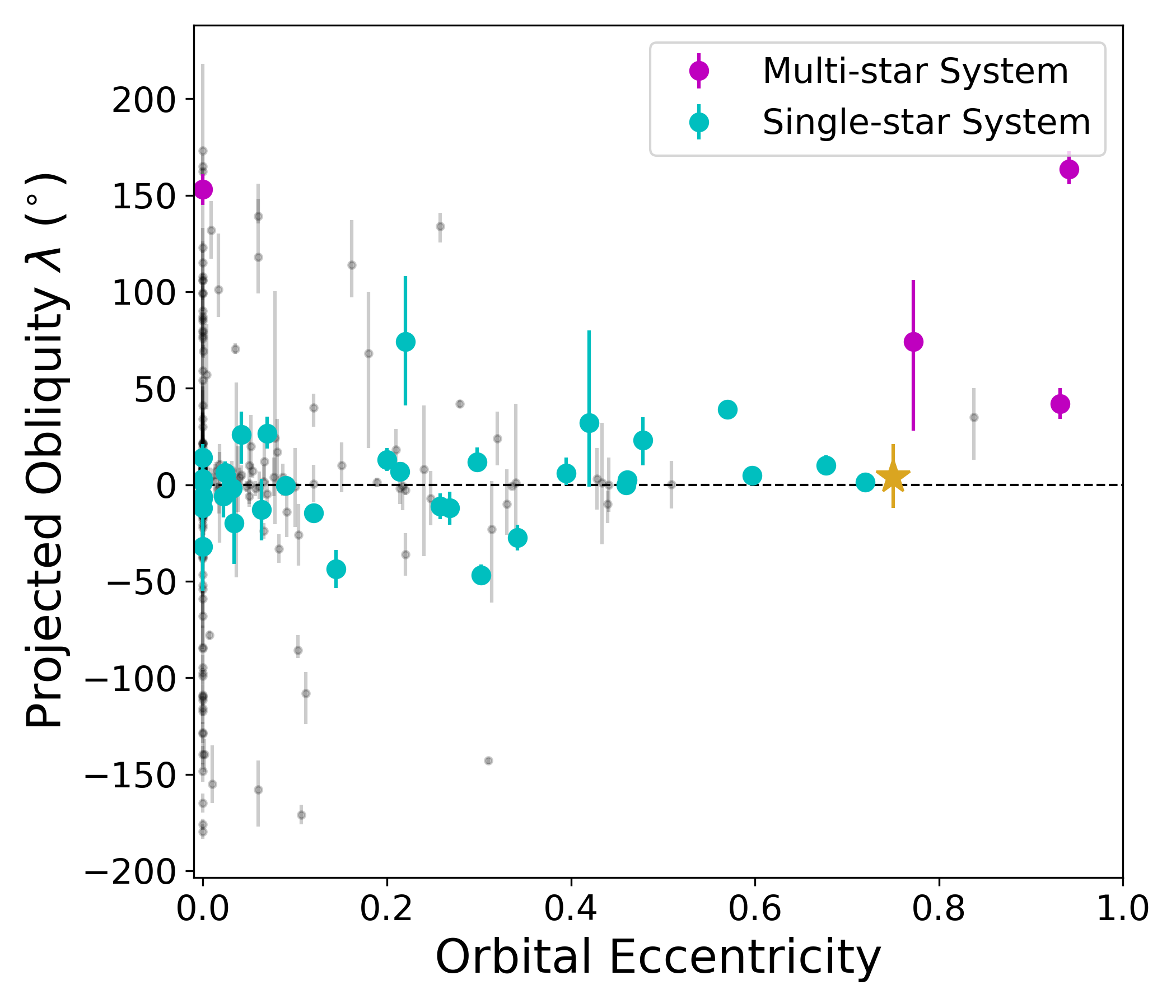}
  \caption{\textit{Left}: Projected obliquities of warm Jupiter systems (10 d $<P<$ 200 d and $R_p > 6\, R_{\oplus}$) as a function of orbital period. Warm Jupiters in single-star and multi-star systems are shown in cyan and magenta, respectively. The single-star TOI-4127 system is shown in gold. All other systems with measured obliquities are shown in black. \textit{Right:} Projected obliquities as a function of orbital eccentricity. System parameters were obtained from TEPCat \citep{2011MNRAS.417.2166S}.}
\label{fig:obl_per_ecc}
\end{figure*}

The joint analysis of the \textit{TESS}, NEID, SOPHIE, and HARPS-N data shows that TOI-4127~b is well-aligned with its host star, with a projected obliquity of $4_{-16}^{+17}\, ^{\mathrm{\circ}}$. This makes it one of the longest-period and most eccentric planets with an obliquity measurement, and the longest-period warm Jupiter on a well-aligned orbit (Figure \ref{fig:obl_per_ecc}). It adds to the growing sample of highly eccentric yet well-aligned warm Jupiters that reside in single-star system compared to the misaligned warm Jupiters in multi-star systems. Of the five most eccentric single-star warm Jupiter systems, three contain warm Jupiters with masses greater than 2 Jupiter masses, including TOI-4127. However, the other two do not have measured masses and thus more measurements are needed to determine the role that planet mass plays in the trends observed. Smaller planets may also follow a similar trend of high eccentricity and low obliquity, as shown by the highly eccentric and possibly aligned sub-Saturn Kepler-1656~b \citep{2024ApJ...971L..40R}, although a much larger sample is needed.

\begin{figure*}
\includegraphics[width=\textwidth]{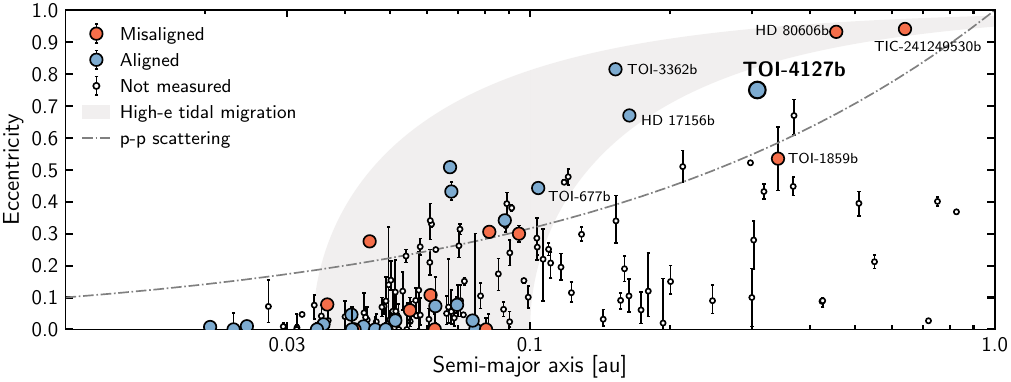}
  \caption{Eccentricity vs. semi-major axis for all confirmed transiting planets between 6--20 Earth radii. Colored circles are planets with measured spin-orbit angles. Blue circles are defined as `aligned' with spin-orbit angles of less than 30 degrees and red circles are defined as `misaligned' with spin-orbit angles greater than 30 degrees. The gray region illustrates planets that are likely undergoing high-eccentricity tidal migration. The upper and lower limits of the track are set by the Roche limit and the tidal circularization timescale \citep{2021ApJ...920L..16D}. The dotted-dashed line presents the theoretical upper limit of eccentricities as a result of planet-planet scattering, assuming a planet with a mass of 0.5 $M_{\rm Jup}$ and a radius of 2 $R_{\rm Jup}$, for illustrative purpose \citep{2014ApJ...786..101P}. Data were compiled from the NASA Exoplanet Archive as of October 30, 2024 \citep{2025PSJ.....6..186C}. Stellar obliquity measurements were obtained from \cite{2024A&A...690A.379K} Table B2 with recent updates.}
\label{fig:sma_ecc}
\end{figure*}

Despite a high orbital eccentricity of 0.75$\pm$0.01, TOI-4127~b is not a hot Jupiter progenitor undergoing high-eccentricity migration unless it is undergoing eccentricity oscillations with intervals of even higher eccentricity. \citet{2023AJ....165..234G} found that the timescale for this planet to migrate and become a hot Jupiter is on the order of 780 Gyr given its current eccentricity. It remains an open question as to how this planet could be so eccentric yet well-aligned given commonly invoked formation and migration mechanisms. Similarly eccentric warm Jupiters like TOI-3362~b and HD 17156~b are closer in to their host stars and, unlike TOI-4127~b, reside in the region of parameter space where high-eccentricity tidal migration can take place (Figure \ref{fig:sma_ecc}). Further out warm Jupiters like HD 80606~b and TIC 241249530~b are more eccentric than TOI-4127~b and reside in known multi-star systems unlike TOI-4127~b.

\subsection{Formation Pathways}

The typical mechanisms invoked for in-situ formation and disk migration cannot explain the system. These mechanisms can explain the low obliquity of TOI-4127, but cannot explain the planet's high eccentricity. Some disk migration mechanisms, e.g., planetary migration in a disk cavity, can excite the eccentricity of a resulting planet, but only up to a value of $\sim$0.4 \citep{2021MNRAS.500.1621D}, far below TOI-4127~b's eccentricity of 0.75. However, more recent dynamical simulations have shown that it is possible for a massive planet in a cavity to obtain high eccentricities solely through resonant interactions with the protoplanetary disk \citep{2023MNRAS.523.2832R, 2024MNRAS.532.3509R}. These interactions excite the eccentricity without misaligning with the orbit and the evolution timescales shorten significantly for more massive planets. The three most eccentric well-aligned warm Jupiters (HD 17156~b, TOI-3362~b, and TOI-4127~b) have masses greater than 2 $M_{Jup}$ and could point towards this mechanism being responsible. However, a larger sample of well-characterized systems is needed to draw any meaningful conclusions. Using the NASA Exoplanet Archive \citep{2025PSJ.....6..186C}, we find that there are eight highly-eccentric ($e > 0.5$) giant planets ($R_p > 6\, R_{\oplus}$) that are amenable for RM observations to measure their obliquties, meaning they orbit stars brighter than V=12.5. These eight planets are TOI-2134~c, TOI-6883~b, TOI-5110~b, TOI-4582~b, TOI-2589~b, TOI-2179~b, Kepler-432~b, and TOI-2338~b. Five planets orbit main sequence FGK stars while the other three orbit evolved stars (two subgiants and one red giant). They range in masses from 0.13 to 5.98 $M_{Jup}$ and in orbital periods from 15 to 111 days. We use Equation 1 from \citet{2018haex.bookE...2T} to estimate the semi-amplitudes of the RM effect, and find that they range from 3 to 24 m s$^{-1}$. While the lower amplitude signals will be challenging, detecting the signals is feasible with current instruments.

Kozai-Lidov oscillations and planet-planet scattering are often cited as the mechanisms responsible for hot Jupiters and their progenitors, failed or otherwise. However, these mechanisms typically result in highly-eccentric, misaligned systems. Kozai-Lidov oscillations could explain this system if we happen to be observing it during a period of low obliquity. In this case, it would be indicative of an additional undetected planetary or stellar companion in the system. Similarly, planet-planet scattering can occasionally result in highly-eccentric well-aligned systems, as has been theorized for HD 17156~b \citep{2009A&A...503..601B}. If there are indeed no additional massive companions in the system, then this scenario would explain the orbit.

Another plausible explanation is that this system is the result of coplanar high-eccentricity migration \citep[CHEM;][]{2015ApJ...805...75P}. This mechanism is thought to be responsible for the well-aligned orbit of the similarly highly-eccentric warm Jupiter TOI-3362~b \citep{2023ApJ...958L..20E}. However, 
TOI-4127~b's larger orbit means its eccentricity is slightly lower than what would be expected for a planet undergoing this process. This can be resolved however, if it is still coupled to the perturber and undergoing eccentricity oscillations. Both this mechanism and some Kozai-Lidov oscillations scenarios require a second giant planet in the system that has yet to be detected. \citet{2023AJ....165..234G} synthesized and examined a sample of perturbers capable of inducing eccentricity oscillations and found that the existing RV data should have found many of such potential planets. Lower mass planets would have avoided detection thus far as would the longest period planets ($\sim$ 10000 d). This sample overlaps significantly with the sample of planets capable of inducing coplanar high-eccentricity migration. A larger fraction of this sample is ruled out, however there are many scenarios where a planet capable of inducing CHEM would not have been detected.

\subsection{Potential Stellar Companions}\label{subsec:comps}

\begin{figure*}
\includegraphics[width=\textwidth]{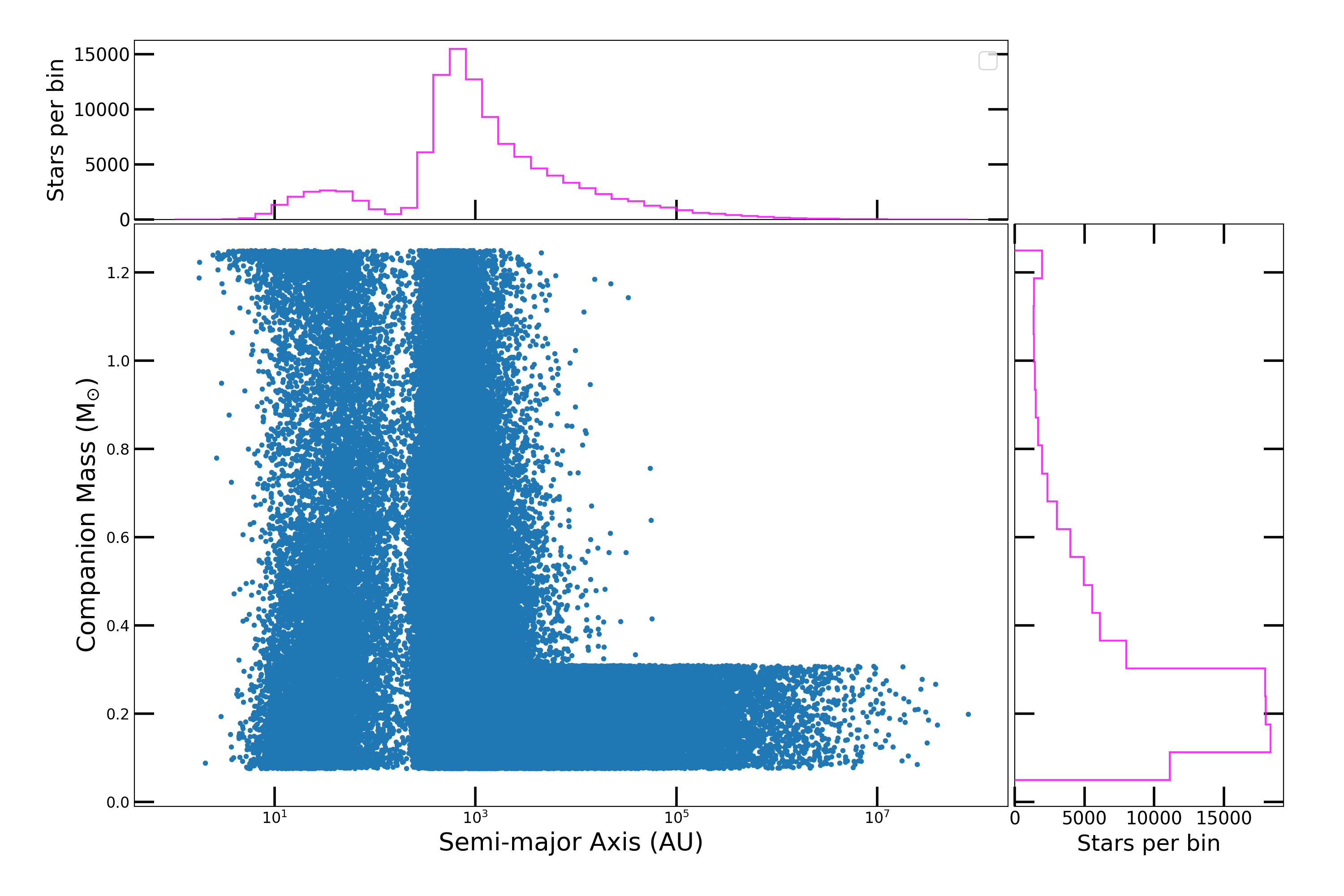}
  \caption{The sample of stars consistent with current observations as determined by MOLUSC. The mass distribution peaks at low masses while the semi-major axis distribution is bimodal.}
\label{fig:molusc}
\end{figure*}

If Kozai-Lidov oscillations are indeed responsible for the orbit of TOI-4127~b, then it could indicate that there is a stellar companion in the system as is the case with HD 80606~b and TIC 241249530~b. While current observations have not led to the detection of a stellar companion, these observations have not probed all of the possible parameter space. While radial velocity measurements have ruled out short-period stellar companions and high-resolution imaging has ruled out bright, distant companions, neither can rule out stellar companions at more intermediate separations. Furthermore, TOI-4127 has a Renormalized Unit Weight Error (RUWE) of 1.27 in Gaia DR3. While RUWE values above 1.4 were considered to be indicative of unresolved binaries in DR2 \citep{2018A&A...616A...2L}, the threshold for EDR3 is lower at 1.25 \citep{2022MNRAS.513.5270P}. Thus, TOI-4127's relatively high RUWE could be the result of an unseen stellar companion, which in turn could be TOI-4127~b's perturber. In order to quantify possible undetected stellar companions, we use Multi-Observational Limits on Unseen Stellar Companions \citep[MOLUSC;][]{2021AJ....162..128W} to generate a sample of potential companions consistent with the combination of the high-resolution imaging, RV data, Gaia astrometry (in the form of the RUWE), and Gaia imaging. Of the 1 million stars we generated using MOLUSC, only $\sim$ 10\% were consistent with the aforementioned data. The sample peaks at low masses, in part due to only faint, low mass stars being allowed at large separations (Figure \ref{fig:molusc}). The semi-major axis is bimodal, with the strongest peak at 1000 AU. Further observations are needed to determine whether there is a stellar companion in this system.  

\section{Conclusion} \label{sec:conclusion}

We analyzed the Rossiter-McLaughlin effect for the highly-eccentric (e=0.75) warm Jupiter system TOI-4127 and found that it is well-aligned, with a sky-projected obliquity of $4_{-16}^{+17}\, ^{\mathrm{\circ}}$. It adds to the growing list of long-period systems with measured obliquities and is one of the most eccentric systems with such a measurement. Multiple mechanisms could explain this highly-eccentric yet well-aligned system, including resonant interactions with the protoplanetary disk, ongoing Kozai-Lidov oscillations, co-planar high-eccentricity migration, and planet-planet scattering.
Kozai-Lidov oscillations and co-planar high-eccentricity migration both require additional companions in the system that have yet to be detected. A distant massive planet or stellar companions would favor Kozai-Lidov oscillations, while a closer-in massive planet would favor co-planar high-eccentricity migration and a lack of any massive companions would favor a scattering event. Additional RV measurements and future astrometric measurements from upcoming Gaia data releases will be able to determine whether there are additional planets or a stellar companion in this system that can help explain the orbit of TOI-4127\,b.

\begin{acknowledgements}
We thank the anonymous referee for their helpful comments, which have helped improve the paper.

We thank Xian-Yu Wang for his assistance with the modified version of \texttt{allesfitter}.

This research has made use of the NASA Exoplanet Archive, which is operated by the California Institute of Technology, under contract with the National Aeronautics and Space Administration under the Exoplanet Exploration Program.

D.D. acknowledges support from the NASA Exoplanet Research Program grant 18-2XRP18\_2-0136.

The Flatiron Institute is a division of the Simons foundation.
\end{acknowledgements}

%

\bibliographystyle{aa}
\bibliography{4127rm.bib}







   
  



\begin{appendix}




\section{HARPS-N Radial Velocities}

In this appendix, we present the values of the HARPS-N radial velocities with uncertainties.

\begin{table}[!h]
\caption{HARPS-N RVs}
\label{table:harpsn_vals}
\centering
\begin{tabular}{l l l}
\hline\hline
BJD-2400000 & RV [m s$^{-1}$] & Instrument \\
\hline
60272.5458 & -36826.10 $\pm$ 6.88 & HARPS-N \\
60385.3467 & -36819.82 $\pm$ 7.10 & HARPS-N \\
60385.3614 & -36820.39 $\pm$ 8.62 & HARPS-N \\
60385.3752 & -36825.41 $\pm$ 5.86 & HARPS-N \\
60385.3889 & -36819.83 $\pm$ 5.74 & HARPS-N \\
60385.4030 & -36826.38 $\pm$ 5.85 & HARPS-N \\
60385.4171 & -36832.40 $\pm$ 6.60 & HARPS-N \\
60385.4327 & -36797.55 $\pm$ 6.93 & HARPS-N \\
60385.4457 & -36798.85 $\pm$ 7.20 & HARPS-N \\
60385.4601 & -36799.48 $\pm$ 7.28 & HARPS-N \\
60385.4738 & -36818.68 $\pm$ 6.88 & HARPS-N \\
60385.4880 & -36832.83 $\pm$ 6.81 & HARPS-N \\
60385.5027 & -36857.17 $\pm$ 6.96 & HARPS-N \\
60385.5169 & -36866.24 $\pm$ 6.83 & HARPS-N \\
60385.5309 & -36880.35 $\pm$ 6.39 & HARPS-N \\
60385.5451 & -36881.75 $\pm$ 7.32 & HARPS-N \\
\hline
\end{tabular}
\end{table}

\section{\texttt{allesfitter} Corner Plots}

In this appendix, we show the corner plots from our \texttt{allesfitter} fit.

\begin{figure*}[!h]
    \centering
    \includegraphics[width=\textwidth]{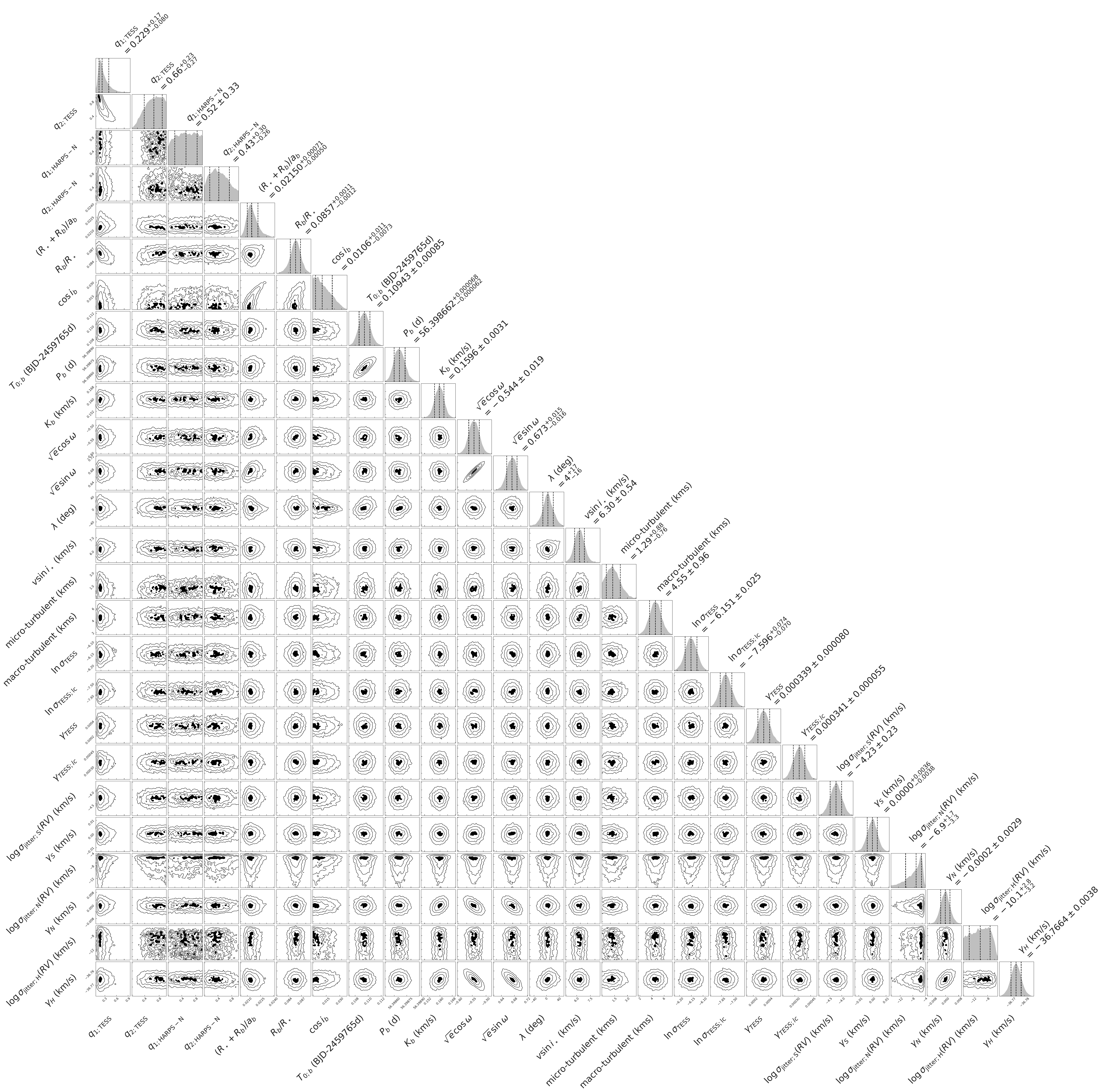}
    \caption{Corner plots of modeled parameters obtained from \texttt{allesfitter}.}
    \label{fig:corner_fitted}
\end{figure*}

\begin{figure*}[!h]
    \centering
    \includegraphics[width=\textwidth]{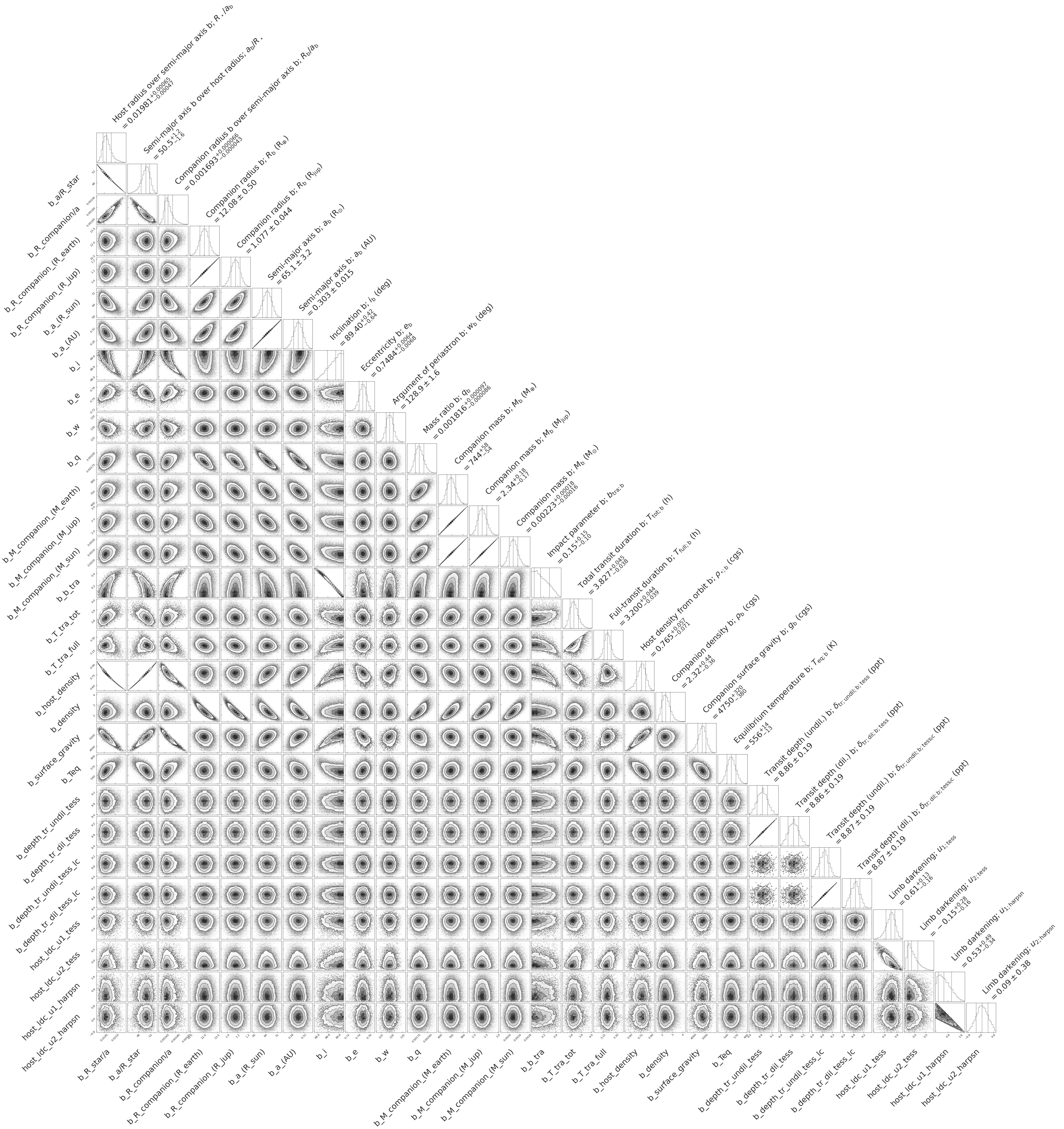}
    \caption{Corner plots of derived parameters obtained from \texttt{allesfitter}.}
    \label{fig:corner_derived}
\end{figure*}

\end{appendix}
\end{document}